\documentclass[twoside]{article}
\usepackage{fleqn,espcrc2,graphicx}
\title{Measurement of Neutron Background at the Pyh\"asalmi mine for CUPP Project, Finland}

\author{J.~N.~Abdurashitov$^{\rm a}$\thanks{Corresponding author: jna@al20.inr.troitsk.ru}, V.~N.~Gavrin$^{\rm a}$, V.~L.~Matushko$^{\rm a}$, A.~A.~Shikhin$^{\rm a}$, V.~E.~Yants\address{Institute for Nuclear Research, Russian Academy of Sciences, Moscow, 117312, Russia}, \\ J.~Peltoniemi$^{\rm b}$, T.~Ker\"anen\address{University of Oulu and CUPP project, Finland}}

\begin{document}

\begin{abstract}

A natural neutron flux is one of significant kind of background in high-sensitive underground experiments. Therefore, when scheduling a delicate underground measurements one needs to measure neutron background. Deep underground the most significant source of neutrons are the U-Th natural radioactive chains giving a fission spectrum with the temperature of 2--3~MeV. Another source is the U-Th~($\alpha$,n) reactions on light nuclei of mine rock giving neutrons with different spectra in the 1--15~MeV energy region. Normal basalt mine rocks contain $10^{-6}$~g/g of $^{238}$U and less. Deep underground those rocks produce natural neutron fluxes of $10^{-7}$--$10^{-6}$~cm$^{-2}$s$^{-1}$ above 1~MeV. To measure such a background one needs a special techniques.

In the Institute for Nuclear Research, Moscow, the neutron spectrometer was developed and built which is sensitive to such a low neutron fluxes. At the end of 2001 the collection of neutron data at the Pyh\"asalmi mine was started for the CUPP project. During 2002 the background and rough energy spectra of neutron at underground levels 410, 660, 990 and 1410~m were measured. The result of the measurement of the neutron background at different levels of the Pyh\"asalmi mine is presented and discussed. Data analysis is performed in different energy ranges from thermal neutrons up to 25~MeV and above.

\end{abstract}

\maketitle

\section{Introduction}

When developing underground physics laboratories it is need to take into account the factors which contribute in background environment. One such factor is neutrons which are produced by natural radioactive U-Th chains in surrounding rocks. At depths which are high enough (2000~m.w.e. and more) the intensity of cosmic ray muons decreases down to $\sim$10$^{-7}$~cm$^{-2}$s$^{-1}$ which is comparable to the background neutron flux from crystalline rocks. Thus, at high depths a relative contribution of neutrons to background of physics experiments may appear to be considerable. Estimation of this contribution into background can be based on calculation of flux using the measured concentrations of U-Th and chemical composition of surrounding mine rocks. Direct measurement of the spectrum and flux of background neutrons is important too. However, a low neutron flux is very difficult to be measured.

Several research groups have investigated the neutron background at different underground laboratories~\cite{chaz1,arn2,alex3}. Some of them used $^6$Li-dopped liquid scintillator technique~\cite{arn2} and others used in addition a pulse shape discrimination technique~\cite{chaz1}. The fluxes of $\sim$10$^{-6}$~cm$^{-2}$s$^{-1}$ in the energy range $>$2~MeV were measured there. Sometimes a radiochemical technique was used there~\cite{gav4}.

In the Institute for Nuclear Research, Moscow, the neutron spectrometer was developed and built which is sensitive to such a low neutron fluxes. The measurements of fast neutron flux in several underground rooms of Baksan Neutrino Observatory (BNO), SAGE~\cite{abd5} and DULB~\cite{abd6}, have been performed with the spectrometer. These rooms are located under Mt.~Andyrchy (Northen Caucasus Mountains, Russia) in a tunnel that penetrates 4.5~km into the mountain, at a depth of 4700 and 4900 meters of water equivalent accordingly. The fluxes of $\sim$10$^{-7}$~cm$^{-2}$s$^{-1}$ in the energy range $>$1.5~MeV were measured there successfully.

The series of neutron flux measurement were carried out according to agreement for a joint Russian-Finnish scientific research programme in 2000--2002 between the Institute for Nuclear Research of the Russian Academy of Sciences (project SAGE) and The Oulu Southern Institute of the University of Oulu (project CUPP, http://www.cupp.oulu.fi) at underground facilities of the Pyh\"asalmi mine, town Pyh\"aj\"arvi, Finland. We present the results here.

\section{Detector}

\subsection{Design}

\begin{figure}[t]
\begin{center}
\includegraphics[scale=0.25]{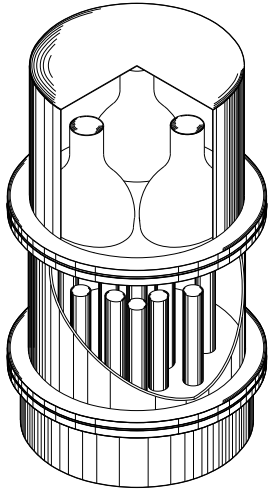}
\end{center}
\caption{General view of the detector}
\label{det}
\end{figure}

The detection part of the spectrometer (detector) consists of a liquid organic scintillator viewed by three photomultipliers (PMT) and  nineteen proportional counters filled with $^3$He (neutron counters --- NC) and distributed uniformly over the scintillator. Figure~\ref{det} shows a general view of the detector. It was described initially in~\cite{abd7}.

The housing of the detector is made of stainless steel. The space of cylinder tank (diameter of 360$\times$360~mm long) available for the scintillator is about 30~l. To enhance light collection the wall and bottom of the scintillator tank are coated with a fluoroplastic. Each counter in the tank is coated with aluminium foil too. The illuminator is made of organic glass and is covered by fluoroplastic plate with windows intended to improve the light collection and to secure the phototubes. The external surface of the tank is 6267~cm$^2$.

\subsection{Principle of operation}

Fast neutrons ($>$1~MeV) entering the scintillator are decelerated down to a thermal energy and diffuse in the detector until they are either captured in a neutron counter or captured by scintillator protons or leave the detector. The intensity of light flash combined from recoil protons which are produced during neutron thermalization is on average proportional to the initial neutron energy. A portion of thermalized neutrons are captured by $^3$He nuclei in the neutron counters, which emit charged particles via the reaction $^3$He(n,p)t, E$_p$=574~keV, E$_t$=191~keV. Thus, a signature of such event is a light flash in the scintillator followed by capture in the neutron counter after some time of delay. This delay is conditioned by the mean lifetime of thermalized neutrons inside the detector and is determined mainly by the detector design. For this particular detector T$_{1/2}$ is $\sim$50~$\mu$s. This technique allows one to suppress the natural background of $\gamma$-rays significantly, by several orders of magnitude.

\subsection{Acqusition system}

\begin{figure}[t]
\includegraphics[width=75mm]{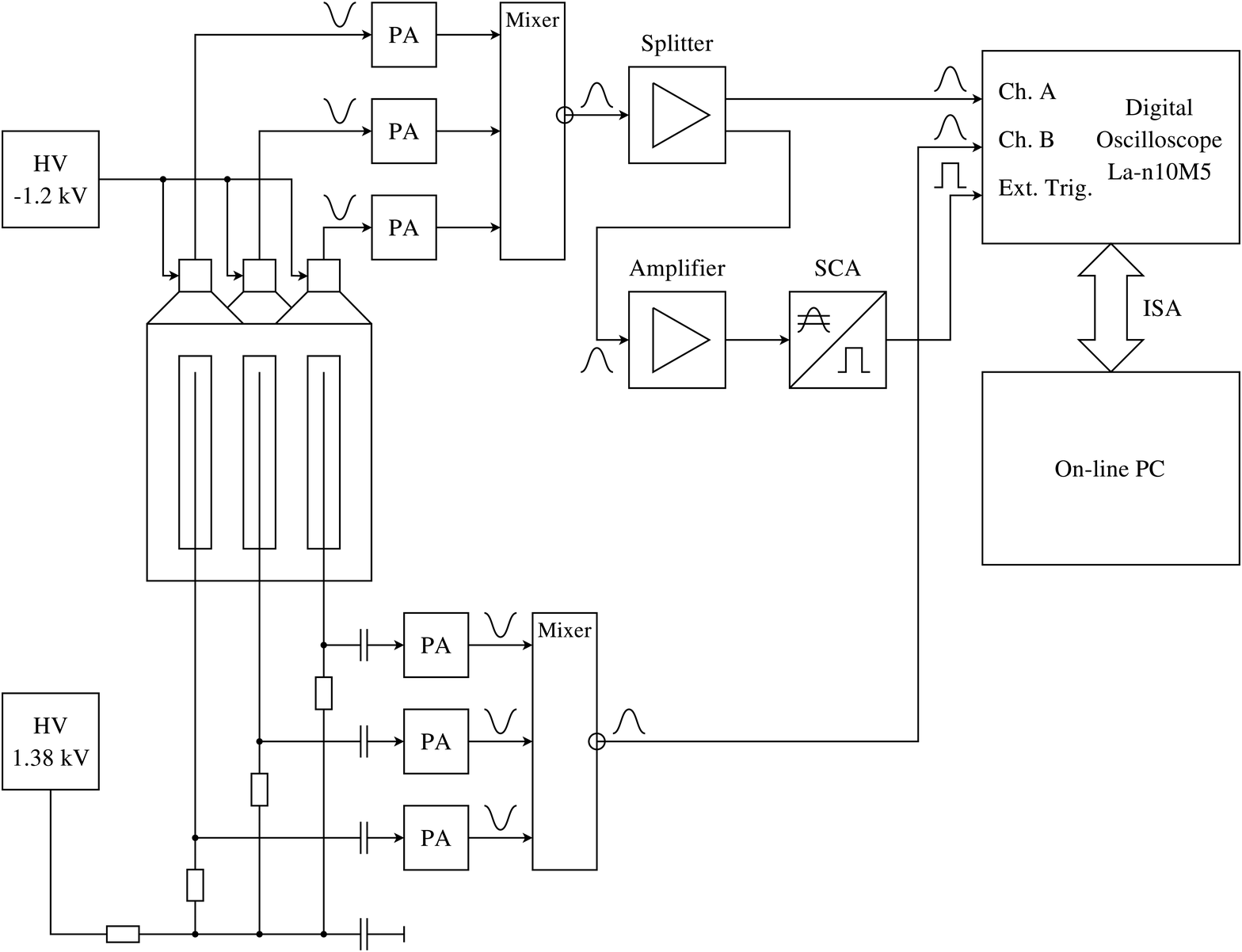}
\caption{The functional diagram of a data acquisition system}
\label{syst}
\end{figure}

The functional diagram of the data acquisition system is shown on Figure~\ref{syst}. In order to simplify the apparatus design the signals from all PMTs and NCs are mixed into two independent channels called "PMT channel" and "NC channel" respectively. The circuit design of the low-level signal electronics are selected taking into account an optimal signal-to-noise ratio. To reduce the noise, the main amplification of initial signals is accomplished with the preamplifiers before joining.

At the PMT channel a negative signal from anode of the PMTs is entered to the input of preamplifier (PA). A continuously adjusted PA gain ensures the operation of all three PMTs from a single high-voltage power supply. The PMT signal from PA output is multiplexed by a fast inverting linear mixer and then is distributed on two directions. One branch entered directly to the input of first channel of digital oscilloscope (DO). The other one is entered to the input of combined unit of amplifier and single-channel analyzer (SCA). The positive TTL-specified signal from SCA output triggers the DO through an external trigger input at the "Spectrum Acquisition Mode" (see below) which is used in calibration procedures, control of the system operation stability and background rate monitoring during acquisition.

In the NC channel, a high voltage of positive polarity from a single source is applied through high-voltage isolating resistors to the anodes of helium counters. The anode generates signal of negative polarity which is fed through high-voltage separating capacitors to the input of PA. The spread of the counter gas amplification factor is compensated by adjusting the PA gain during calibration. For convenient channel tuning each of counter PA (similarly to the PMT channel) can be switched off independently by turning off the PA supply voltages. The signal of neutron counter PA outputs are multiplexed in a single output fast inverting linear mixer and subsequently enter directly to the input of the second channel of DO. A signal from NC channel triggers the data acquisition system at the "Pulse Acquisition Mode" which is usually used for real background measurements.

The full waveform of event in the PMT and NC channels is recorded independently by means of the two-channel PC/AT-interfaced digital oscilloscope LA-n10M5 manufactured by "Rudnev-Shilyaev" Co., Moscow, Russia. It's performance data are shown in Table~\ref{table1}. The frame is recorded in selected time interval including periods before and after the trigger which are called prehistory and history accordingly. This time interval can be adjusted in the acquisition algorithm over the wide range (2~$\mu$s to 2~ms). Usually the trigger, that is a capture in $^3$He counter, is put in the middle of scope frame ($-$160 to 160~$\mu$s). Thus, the prompt PMT event being correlated with the $^3$He count should appear before it in the interval $-$160 to 0~$\mu$s and uncorrelated one may appear elsewhere with the flat probability over the whole frame.

\begin{table*}[t]
\caption{Main performance data of the digital oscilloscope LA-n10M5}\label{table1}
Parameters are presented as for 2-channel (1-channel) mode.%

\begin{tabular}{ll}
\hline
Independent A/D converter & 2 \\
Resolution of A/D conversion & 8 bit \\
Sample frequency range & 3.125~kHz--50~MHz~(6.104~kHz--100~MHz)\\
Internal RAM & 256~kB, 128~kB per channel \\
Input sensitivity, V & $\pm$1, $\pm$0.5, $\pm$0.2, $\pm$0.1 \\
Input impedance & 1 M$\Omega$ and 15 pF \\
S/N ratio & 47~dB \\ \hline
\end{tabular}
\end{table*}

The acquisition system operates under control of code NOSC especially designed for this purpose. It is written in C and is executed in real-time mode MS-DOS 6.22. Mainly, it is a virtual oscilloscope recording frame by frame in "Pulse Acquisition Mode". In addition, it is able to collect a pulse height histogram in "Spectrum Acquisition Mode".

\subsection{Sensitive elements}

The scintillator "Protva" is developed and manufactured by M.~Konoplya, IHEP, Protvino, Russia. It is similar to NE-213 and has the density of 0.84~g/cm$^3$, the hydrogen concentration of 136~g/kg, the flash duration of 3~ns, the light yield at least 40\% of antracene and the ignition temperature of 80$^{\circ}$C. The scintillator is viewed with three phototubes PMT Model~173, manufactured by "Ekran", Novosibirsk, Russia. The photocathode of PMT is 160~mm in diameter, so three tubes cover about 70\% of top side of the tank. The PMT Model~173 isn't designed for fast timing applications, so the rise time is rather poor --- about 100 ns. The preamplifiers of signals from PMTs and the signal mixer are placed at the top of the detector and are covered with an optically isolated cap.

Neutron counters are commercially available SNM-18, Russia. Nineteen counters (diameter of 30$\times$300~mm long) filled with a $^3$He$+$4\%Ar mixture at a pressure of 4~atm are mounted in sealed wells in the internal space of the tank. The width of current signal is about 1$\div$4~$\mu$s. The anode contacts of the counters are terminated under the bottom of the tank. The preamplifiers of NCs and the signal mixer are placed at the bottom of the detector and are also covered with a cap.

\subsection{Efficiency}

The efficiency was simulated in two stages. First, the history of the neutron with the energy $E_0$ initially incident on the surface of the detector was traced individually taking into account all kinds of an energy loss. A special code DGEOM was developed there to trace the neutron. It traces the neutron down to 50~keV fixing those recoil protons which are able to produce a detectable light flash. It fixes those events where the neutron leaves the detector out also. As a result one finds a geometrical efficiency $g(E_0)$, i.e. the probability of the neutron to be trapped in the detector depending on initial energy $E_0$. The DGEOM does a correction of probability of incident neutrons by a cosine of falling angle, so one can use a geometrical surface of the detector in a data analysis.

Second, a standard code MORSE was used to simulate the efficiency $\varepsilon_{th}$ of capture of neutron in the detector. It's assumed that neutrons are burned in the detector uniformly with the energy 50~keV --- a final energy of previous stage. It is resulted in $\varepsilon_{th}=(20\pm1)$\%. It was checked with Pu-Be source and found to be agreed within 10\% uncertainty.

\begin{figure}[t]
\includegraphics[width=75mm]{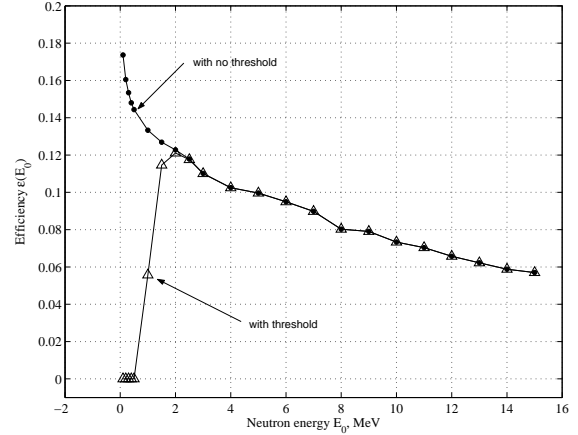}
\caption{Efficiency plot}
\label{effic}
\end{figure}

Last, the efficiency of the detector to trap the neutron and to measure it's energy is found as
\begin{equation}
\varepsilon(E_0)=\varepsilon_{th}\times(E_0).
\end{equation}
It is shown on the Figure~\ref{effic}.

\section{Calibration of energy scale and time of delay}

\subsection{NC channel}

\begin{figure}[t]
\includegraphics[width=75mm]{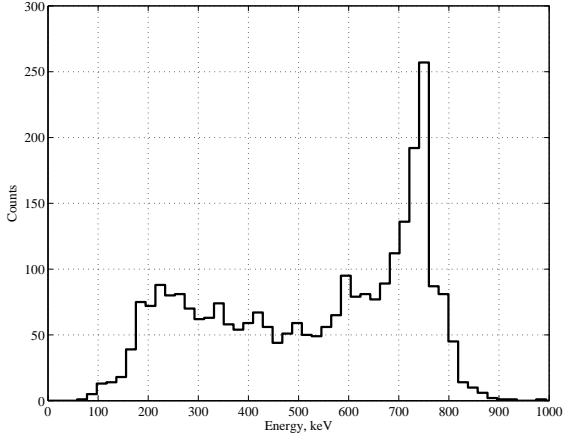}
\caption{The pulse height spectrum of mixed helium counters events when irradiated with Pu-Be neutron source}
\label{pube_he3}
\end{figure}

The Pu-Be neutron source with intensity of $2\times10^3$~s$^{-1}$ was used to irradiate the detector keeping in mind an unification of individual spectrum of each $^3$He proportional counter. The pulse height spectrum of mixed nineteen counters is shown on Figure~\ref{pube_he3}. Two peaks are clearly expressed there: first one is located near 191~keV of tritium energy and second one is located near 760~keV of both tritium and proton energies. The neutron window of acceptance is choosen to be 170--870~keV, that are 35--180~channels of ADC in $\pm100$~mV range. The window accepts 90\% of total captured neutrons. The 760~keV peak is found at $157\pm2$~channel of ADC, so the weight of one ADC channel in $\pm100$~mV range is 4.84~keV at 1270~V of high voltage supplied. The uncertainty of the scale is estimated to be $\pm1$\%.

\subsection{PMT channel}

\begin{figure}[t]
\includegraphics[width=75mm]{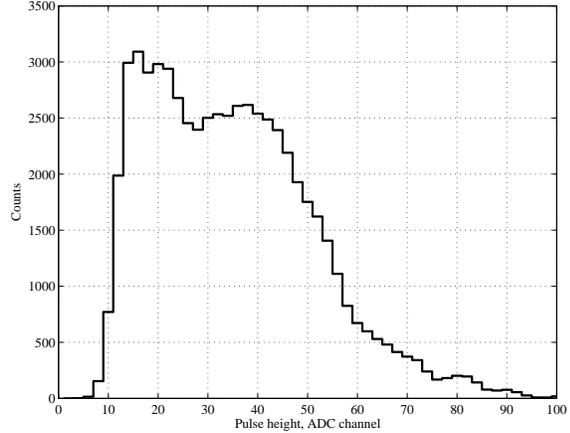}
\caption{The pulse height spectrum of mixed PMT events when irradiated with $^{60}$Co $\gamma$-source}
\label{co60}
\end{figure}

A $^{60}$Co $\gamma$-source (1.17 and 1.33~MeV lines) was used to calibrate the PMT channel. Due to low Z the $\gamma$-ray absorption ability of the scintillator is rather weak $(d_{1/2}\sim12~cm^{-1}$ for $E_g=1$~MeV). In the frame of particalur design of the detector one can't observe the peak of full absorption of $^{60}$Co. Therefore, a Compton edge location was used to calibrate the PMT channel. The detecor response on $\gamma$-rays was simulated there, and one found that the peak of Compton edge should appear at 1.06~MeV. Figure~\ref{co60} shows a pulse height spectrum of PMT signals when irradiated with $^{60}$Co. It resulted in the weight of one ADC channel to be equal to 26.5~keV at 950~V of high voltage supplied. One should note here that there is some nonuniformity of light collection. The same light flashed both at the top and at the bottom of the tank will result in different PMT signals. The difference was measured to be 20\%, so it is the main contribution in the uncertainty. Thus, the uncertainty of an electron equivalent energy scale of the PMT channel is estimated to be $\pm10$\%.

\begin{figure}[t]
\includegraphics[width=75mm]{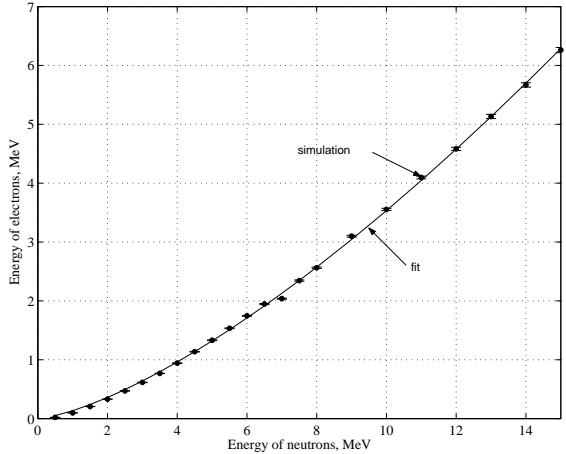}
\caption{The electron v. neutron energy scale for the NE-213 scintillator}
\label{escale}
\end{figure}

The relation between electron and neutron energy scales was obtained as a secondary result of the efficiency simulation (see Figure~\ref{escale}) keeping in mind a specific light yield of proton in the scintillator. This light yield is provided in~\cite{scin8} and is expressed as
\begin{equation}
E_\gamma=0.95E_p-8(1-\exp(-0.1E_p^{0.9})),
\end{equation}
where $E_\gamma$ and $E_p$ are in MeV. It results in negative light yield for $E_p<0.2$~MeV, so we use simple quadratic form for the $E_p<0.5$~MeV energy range. It was found that the light flash produced by recoil protons scattered by neutron with $E_n$ is on average equivalent to the light flash emitted by electron with $E_e$, which is expressed as
\begin{equation}
E_e=0.134E_n^{1.421},
\end{equation}
where $E_e$ and $E_n$ are in MeV.

Finally, the energy of neutrons expressed in ADC channels and electron equivalent MeV is presented in Table~\ref{table2}.

\begin{table*}[t]
\caption{Intervals of neutron energy expressed in different units for two modes of measurement}\label{table2}
\begin{tabular}{llllllll}
\hline
Mode&Units&\multicolumn{6}{l}{Neutron energy interval}\\
&MeV,~neutron~scale&0--1.5&1.6--3&3--6&6--12&12--25&$>$25\\ \hline
LOW&MeV,~electron~scale&0--0.25&0.26--0.62&0.63--1.65&1.66--4.43&--&--\\
&ADC~chan.,~$\pm$100~mV&0--9&10--24&25--63&64--168&--&--\\
HIGH&MeV,~electron~scale&--&--&--&--&4.4--12.6&$>$12.6\\
&ADC~chan.,~$\pm$200~mV&--&--&--&--&84--255&256\\ \hline
\end{tabular}
\end{table*}

\subsection{Correlated signals}

\begin{figure}[t]
\includegraphics[width=75mm]{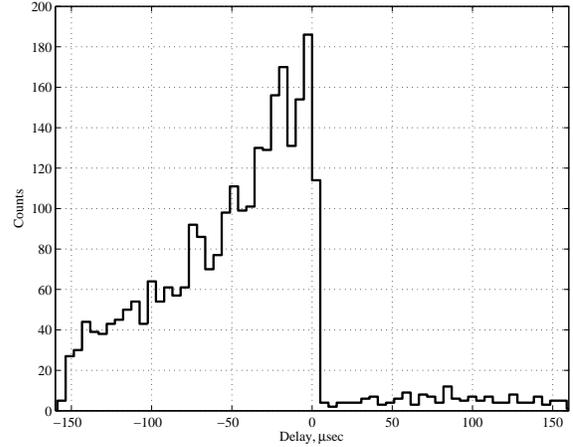}
\caption{The time of delay distribution of fast neutron events (Pu-Be source)}
\label{delay}
\end{figure}

The correlated events from the detector irradiated with Pu-Be neutron source are used to find a specific lifetime of thermalized neutrons before capture in helium counters. The delay time distribution for the neutron source is presented on Figure~\ref{delay}. The best fit of lifetime is $T_{1/2}=50\pm2$~$\mu$s.

\section{Data acquisition}

We started the collection of neutron data at the Pyh\"asalmi mine at the end of 2001. During 2002 we were measuring the background and rough spectra of neutron at underground levels 410, 660, 990 and 1410 m. In the Table~\ref{table3} one can find history of measurements. The files have being recorded 55~min in "Pulse Mode" and 5~min in "Spectrum Mode" each hour.

There are two modes of measurement --- LOW and HIGH (see Table~\ref{table2}) which differs mainly by range of PMT ADC ($\pm100$ or $\pm200$~mV). The LOW mode is used to find the neutron flux in the 0--12~MeV range as well as the HIGH one is used for $>$12~MeV energies.

\begin{table}[t]
\caption{Chronology of the measurements}
\label{table3}
"LOW" energy range means 1.5--12~MeV, \\
"HIGH" means $>$12~MeV.%

\begin{tabular}{llll}
\hline
Level&Mode&Data files&Date\\ \hline
1410&LOW&J112212&21 Dec--03 Jan \\
&&J201032&03 Jan--11 Jan \\ 
440&LOW&J202130&13 Feb--26 Feb \\
&&J202270&27 Feb--06 Mar \\
&HIGH&J203060&06 Mar--25 Mar \\
&&J203270&27 Mar--15 Apr \\
&&J204150&15 Apr--06 May \\
990&LOW&J205070&07 May--27 May \\
&HIGH&J206110&11 Jun--02 Jul \\
660&LOW&J207020&02 Jul--29 Jul \\
&HIGH&J207290&29 Jul--13 Aug \\
1410&HIGH&J209131&13 Sep--27 Sep \\ \hline
\end{tabular}
\end{table}

The passive shield made from standard lead sheets ($1000\times500\times10$~mm) in four layers was used there. It is used to suppress by several times a background $\gamma$-rays producing accidental coincidences with a negligible distorting of the neutron spectrum. The $\gamma$-ray counting rate under the shield varied in 400--600~s$^{-1}$ range above 100~keV over all levels of the mine.

\section{Events characterization}

\subsection{Normal events}

Normal events are reasoned by particles producing both a light flash in the scintillator and current signal in the proportional counter. One significant criterium is a shape of pulses produced by the preamplifiers. Normal PMT event should have the rise time to be $\sim$0.1~$\mu$s which is reasoned by the time of electron collection in the photomultiplier. The fall time is reasoned only due to discharge of correspondent input and output RC circuits of the preamplifier and has width $\sim$30~$\mu$s. Normal NC event has, in contrast, the rise time 1$\div$4~$\mu$s due to internal properties of the proportional counter. Again, the fall time is reasoned only due to discharge of the RC circuits at the input and output of preamplifier and has width $\sim$60~$\mu$s.

\begin{figure}[t]
\includegraphics[width=75mm]{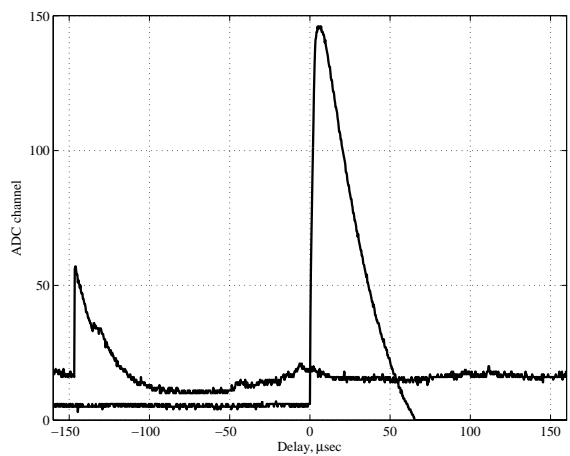}
\caption{Normal event with the PMT signal before. Upper line --- PMT, lower --- NC}
\label{nbefore}
\end{figure}

Another significant criterium is a relative appearance of signals in the frame of the scope. One example of normal event is shown on Figure~\ref{nbefore}. It is a good candidate to be a neutron thermalized in the scintillator with prompt PMT signal and captured in one of the $^3$He counters. Such event is most probably a correlated coincidence. Another source of such events is a decay of Bi and Po radioactive isotopes, which can take place in a helium counter wall. It has been considered as main possible source of an internal background. It imitates an actual neutron event when beta decay of $^{214}$Bi firing the scintillator is followed by delayed ($T_{1/2}=164~\mu$s) $\alpha$-decay of $^{214}$Po in helium counters. The delay time distribution of correlated events was measured in special water shield suppresing the neutron background completely~\cite{abd5}. It was found that the $T_{1/2}$ is very close to 164~$\mu$s confirming thus the origin of the detector internal background.

\begin{figure}[t]
\includegraphics[width=75mm]{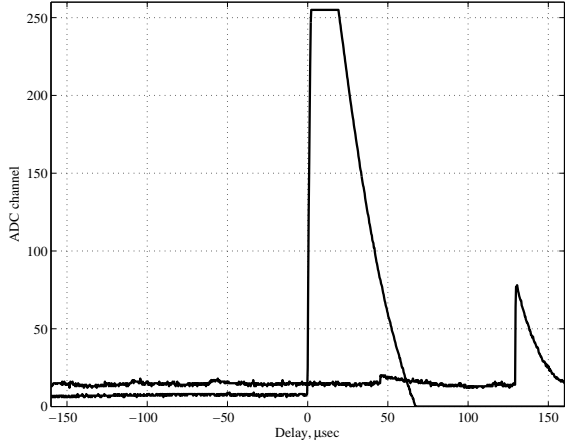}
\caption{Normal event with the PMT signal after. Upper line --- PMT, lower --- NC}
\label{nafter}
\end{figure}

There is some probability for the event to be completely uncorrelated, or an accidental coincidence. The source of accidental coincidence is an independent light flash, produced by a background $\gamma$-ray with simultaneous capture of a background thermal neutron or $\alpha$-particle in helium counter. For those events the PMT signal may appear before as well as after the helium counter signal with the same probability. An example of such event is shown on Figure~\ref{nafter}. The helium counter signal is reasoned most probably by uncorrelated $^{214}$Po decay here because of the amplitude doesn't fit the neutron window of acceptance (that is overflowed). A passive shield suppresses those events significantly.

\begin{figure}[t]
\includegraphics[width=75mm]{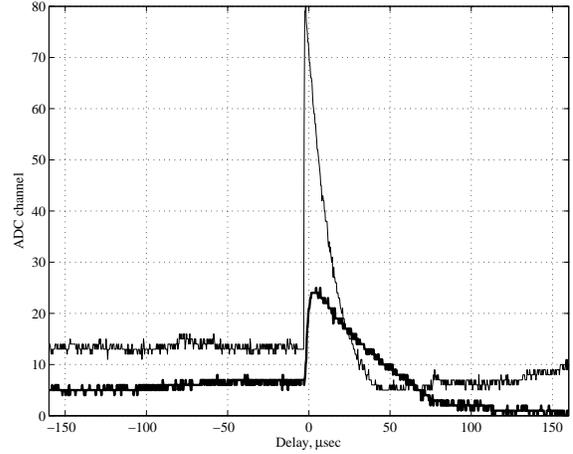}
\caption{Normal simultaneous event. Upper line --- PMT, lower --- NC}
\label{simult}
\end{figure}

Normal event may appear also when a background $\gamma$-ray produces an electron that fires both the scintillator and the helium counter at the same time. It results in the frame recorded where the PMT signal is located at the same time position as the NC signal is. An example of such event is shown on Figure~\ref{simult}.

\subsection{Breakdown, Discharge and Noise \\ events}

\begin{figure}[t]
\includegraphics[width=75mm]{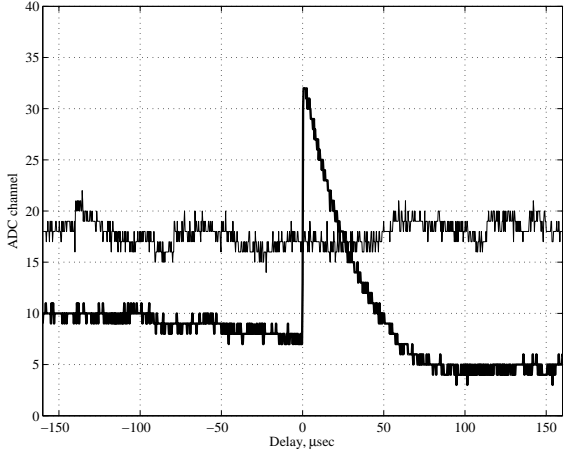}
\caption{NC breakdown event. Upper line --- PMT, lower --- NC}
\label{breakd}
\end{figure}

Sometime a breakdown event in helium counter may appear there. An example of such event is shown on Figure~\ref{breakd}. One can easily note that the rise time of the event is rather fast in comparison to normal event. This is due to that a breakdown is a discharge appearing outside the counter just, for example, through water traces on the surface of counter. Due to variation of humidity in the mine air sometimes one observes high breakdown rate for a long time (several hours or more).

\begin{figure}[t]
\includegraphics[width=75mm]{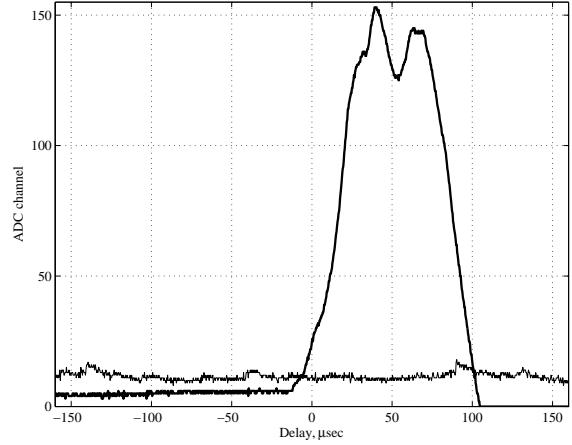}
\caption{NC discharge event. Upper line --- PMT, lower --- NC}
\label{dischrg}
\end{figure}

An origin of discharge events in helium counter is most probably a discharge through air outside the counter. Due to low drift velocity it is rather slow one (see Figure~\ref{dischrg}).

\begin{figure}[t]
\includegraphics[width=75mm]{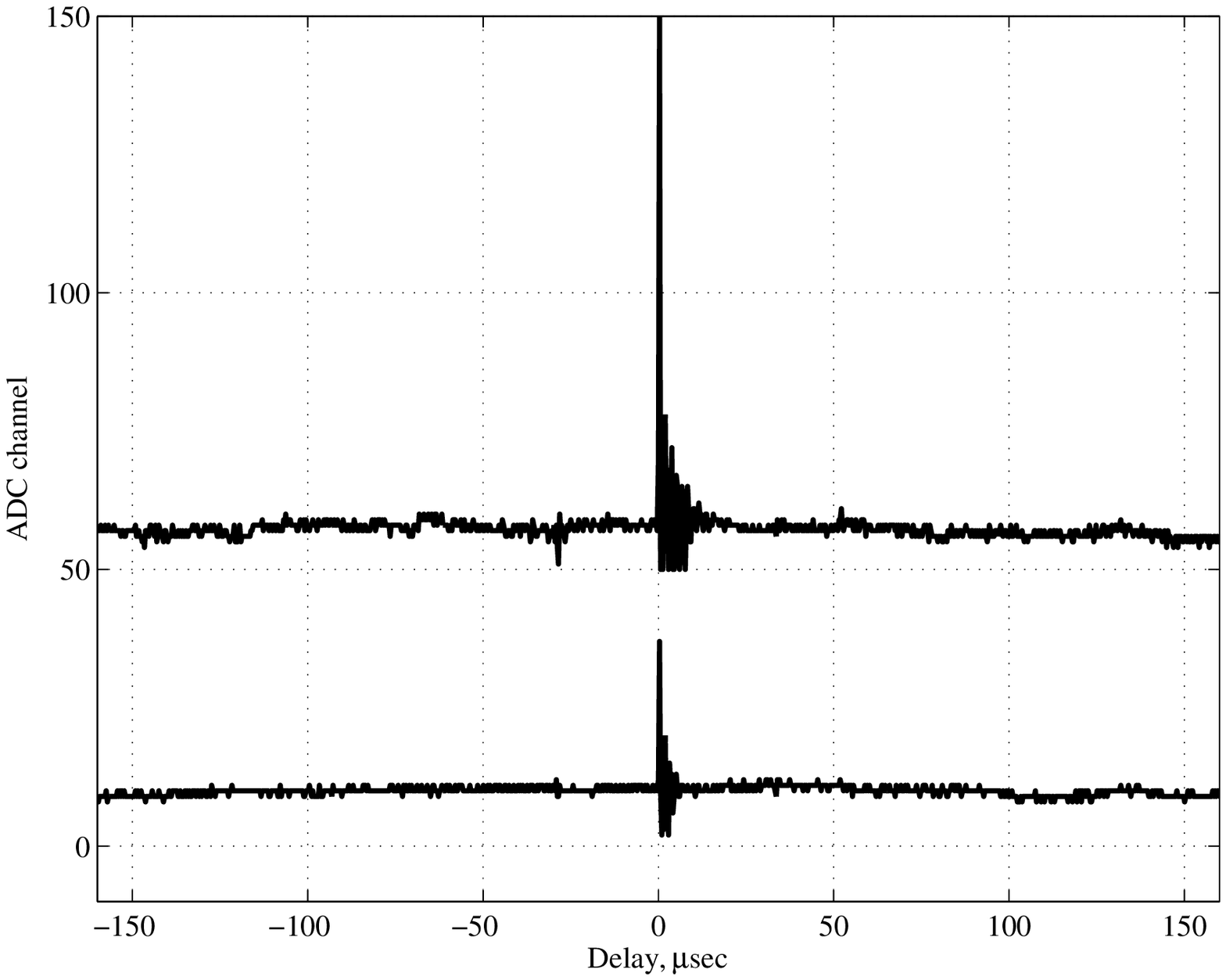}
\caption{NC and PMT noise event. Upper line --- PMT, lower --- NC}
\label{noise}
\end{figure}

Noise events may appear due to high frequency current induced by power turning off/on or some other reasons. An example of the noise event is shown on Figure~\ref{noise}.

\subsection{Discrimination}

It's easy to discriminate completely breakdown, discharge and noise events due to a very specific pulse shape of the signals. It's easy also to reduce a pure accidental coincidence when the PMT signal appears after the NC one. This kind of discrimination is applied to data.

In contrast, for the particular design of the detector it is practically impossible to distinguish a different sources of the individual correlated events. Nevertheless, it becomes quite possible by means of statistical method when one collect different events at the same time of single measurement. So, a pure fast neutron signal is obtained as a remainder of the correlated data substracted with the accidental and Bi-Po background.

\section{Data analyzis}

An immediate result of measurement is a file containing the frames recorded. The analysis of data is proceeded in four stages.

First, one analyses periods of the event rate unstability. Those periods may appear, for example, due to high humidity resulting in a high rate of breakdowns. Such periods are rejected if exist. Then a live time of measurement is calculated taking into account a dead time, a rejected intervals and total number of frames recorded. At this stage a total number of normal single events in helium counters are fixed also in two energy regions --- $T_{nw}$ in neutron window (35--180 ADC~ch, or 170--870~keV) and $T_{aw}$ in $\alpha$-window (181--248 ADC~ch, or 870--1200~keV). The $T_{nw}$ is corrected by a factor 1.1 of the neutron window acceptance. The $T_{aw}$ is corrected to the interval 35--180 ADC~ch by a factor $(248-181)/(180-35)=2.16$. Both $T_{nw}$ and $T_{aw}$ are corrected by a factor 19/(number of active helium counters). Also at this stage a relative uncertainty $U_{es}$ of those numbers appearing due to uncertainty of the enegy scale is estimated. It is obtained as a variation of $T_{nw}$ and $T_{aw}$ when the boundaries of the windows were varied.

Second, four PMT pulse height histograms are collected on the data being remained after rejection. Initially, a threshold of a PMT signal to be accepted is obtained keeping in mind that the amount of double (triple,~$\ldots$) PMT events in a single frame should be less than (5--7)\% of total PMT events. Usually it results in the threshold of 10~ADC~ch corresponding to 265~keV of electron scale, or 1.6~MeV of neutron scale. Then, a simultaneous events time window $STW$ is obtained from the delay time distribution of PMT events coincident with NC events below the neutron window ($<$170~keV). Usually $STW=[-10;6]$, or 16~$\mu$s. Thus, one obtains an accidental $ATW$ and a correlated $CTW$ time windows also (usually $CTW=[-160;-10]~\mu$s, $ATW=[6;160]~\mu$s). After that, the PMT pulse height histograms $PNC$ and $PNA$ are collected for those PMT signals which are coincident with helium counter events in the neutron energy window and in the definite time window ($CTW$ and $ATW$ respectively). Similarly, the $PAC$ and $PAA$ are collected for those PMT signals which are coincident with helium counter events in the $\alpha$-window and in the time windows $CTW$ and $ATW$ respectively. Again, the proper histograms are corrected by the neutron window acceptance, the number of active helium counters and the $\alpha$-neutron windows width difference as well as it done for the $T_{nw}$ and $T_{aw}$.

\begin{figure}[t]
\includegraphics[width=75mm]{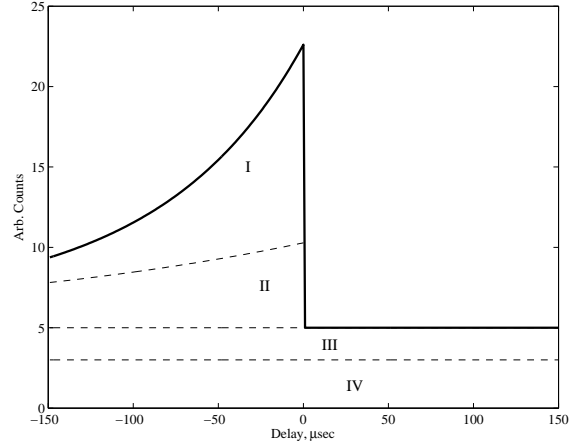}
\caption{The model of distribution of time of delay in real measurement. See text for the labels}
\label{model_bw}
\end{figure}

Third, the pure fast neutron signal is derived from the data. To make it clear one may look at the Figure~\ref{model_bw} where a typical delay time distribution is simulated. It consists of pure fast neutron (region I) and pure Bi-Po (region II) events being coincident in the frame with specific time constants. Those events are grounded at a flat (with no characteristic time constant) distributions of alphas (region III) and thermal neutrons (region IV) accidentally coincident with PMT signals. So, the $PNC$ histogram is a sum of pure fast neutrons, pure Bi-Po events and accidental background (both $\alpha$ and thermal neutrons). The last component is obtained through direct measurement (the $PNA$ histogram). The correlated Bi-Po histogram is obviously obtained as a remainder of the $PAC$ subtracted with the $PAA$. Thus, the fast neutron histogram $FNS$ is obtained as follows:
\begin{equation}
FNS=PNC-PNA-(PAC-PAA).
\end{equation}

This procedure is applied to each particular energy interval (see Table~\ref{table2}) of the spectrum. The uncertainties are statistical which propagate normally and systematic one ($U_{es}$, see above). The result of the procedure is corrected by the factor of the $CTW$ window acceptance 1.38. Then the fast neutron flux is calculated with the efficiency averaged over each particular energy interval.

Last, the slow neutron (i.e., under PMT threshold) signal is obtained as
\begin{equation}
SNS=FNS-(Tnw-Taw),
\end{equation}
each being corrected with proper efficiency. The $FNS$ here is an integral of the PMT pulse height spectrum over all energy intervals.

\section{Example of data}

\begin{figure}[t]
\includegraphics[width=75mm]{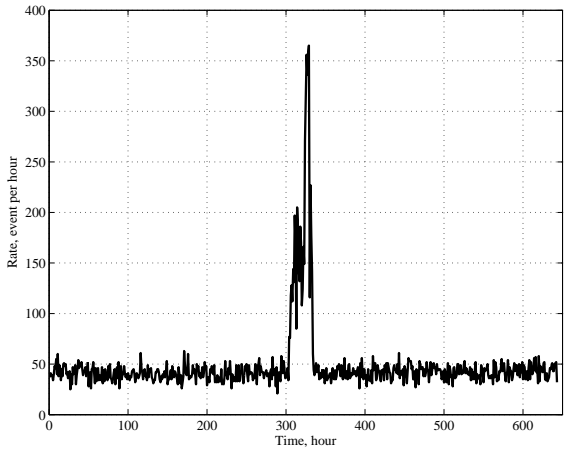}
\caption{$^3$He count rate time behavior at 660~m}
\label{stable}
\end{figure}

Let us consider a typical data measurement made at 660~m in low energy released region. Both PMT and NC channels were recorded in $\pm100$~mV dynamic range of the oscilloscope. This corresponds to [0.21--6.78]~eeMeV (ee --- electron equivalent) and [70--1240]~keV ranges of the PMT and NC channels respectively. The frequency of digitizing was 3.125~kHz with 1024 points of frame (327.68~$\mu$s long). The file J207020.osc was collected for 643~h during 2.07--29.07.2002 and contains 178285 frames. First analysis of the frame rate behavior revealed a period of unstability $336-297=39$~h. Due to breakdowns at the period the frame rate increased drastically --- see Figure~\ref{stable}, so the period was rejected. It resulted in 68085~frames remained after rejection. Each frame recording takes 60~ms, so the dead time appears to be $0.06\times68085/3600=1.13$~h. Thus, the live time is equal to $55\times(643-39)/60-1.13=552.5$~h.

\begin{figure}[t]
\includegraphics[width=75mm]{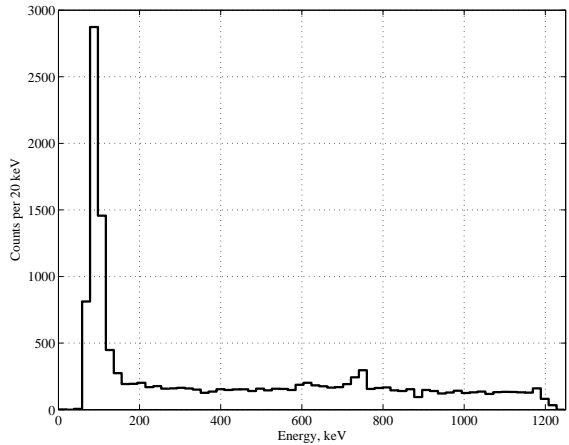}
\caption{Pulse height spectrum of $^3$He counters at 660~m}
\label{hetot}
\end{figure}

For NC normal single events the pulse height histogram is collected there (Figure~\ref{hetot}). One can clearly see the neutron signal expressed there as a typical peak at 760~keV. It is found that the histogram contains $T_{nw}=17781\pm133(stat)\pm136(syst)$ in the neutron window and $T_{aw}=6899\pm83(stat)\pm89(syst)$ in the $\alpha$-window. After corrections by $\alpha$-to-neutron windows 2.164, by active neutron counters $19/18=1.055$ and by neutron window acceptance 1.1 (only $T_{nw}$) and with quadratic sum of uncertainties one obtains $T_{nw}=20646\pm221$ and $T_{aw}=15761\pm734$~events.

\begin{figure}[t]
\includegraphics[width=75mm]{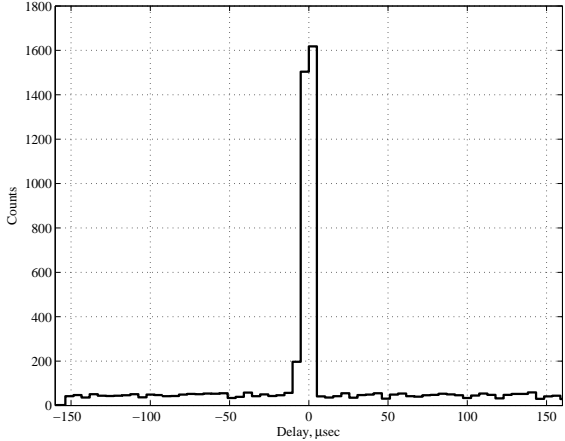}
\caption{The delay time distribution for the simultaneous window at 660~m}
\label{electr_del}
\end{figure}

\begin{figure}[t]
\includegraphics[width=75mm]{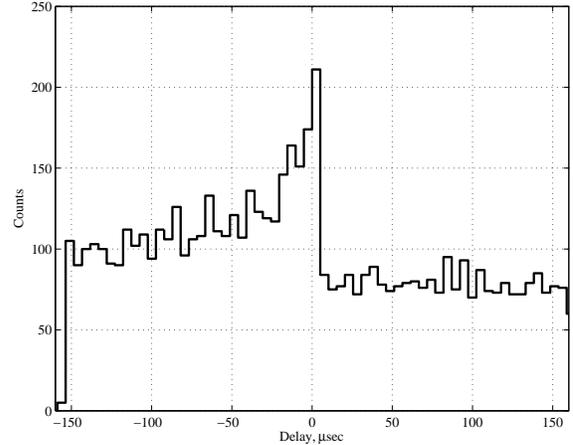}
\caption{The delay time distribution for the neutron window at 660~m}
\label{neutr_del}
\end{figure}

\begin{figure}[t]
\includegraphics[width=75mm]{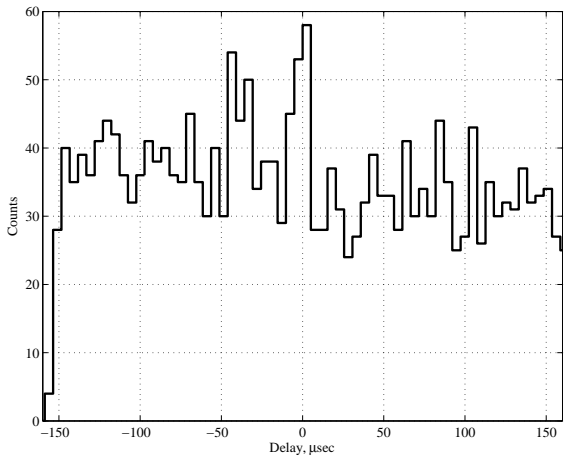}
\caption{The delay time distribution for the $\alpha$-window at 660~m}
\label{alfa_del}
\end{figure}

\begin{figure}[t]
\includegraphics[width=75mm]{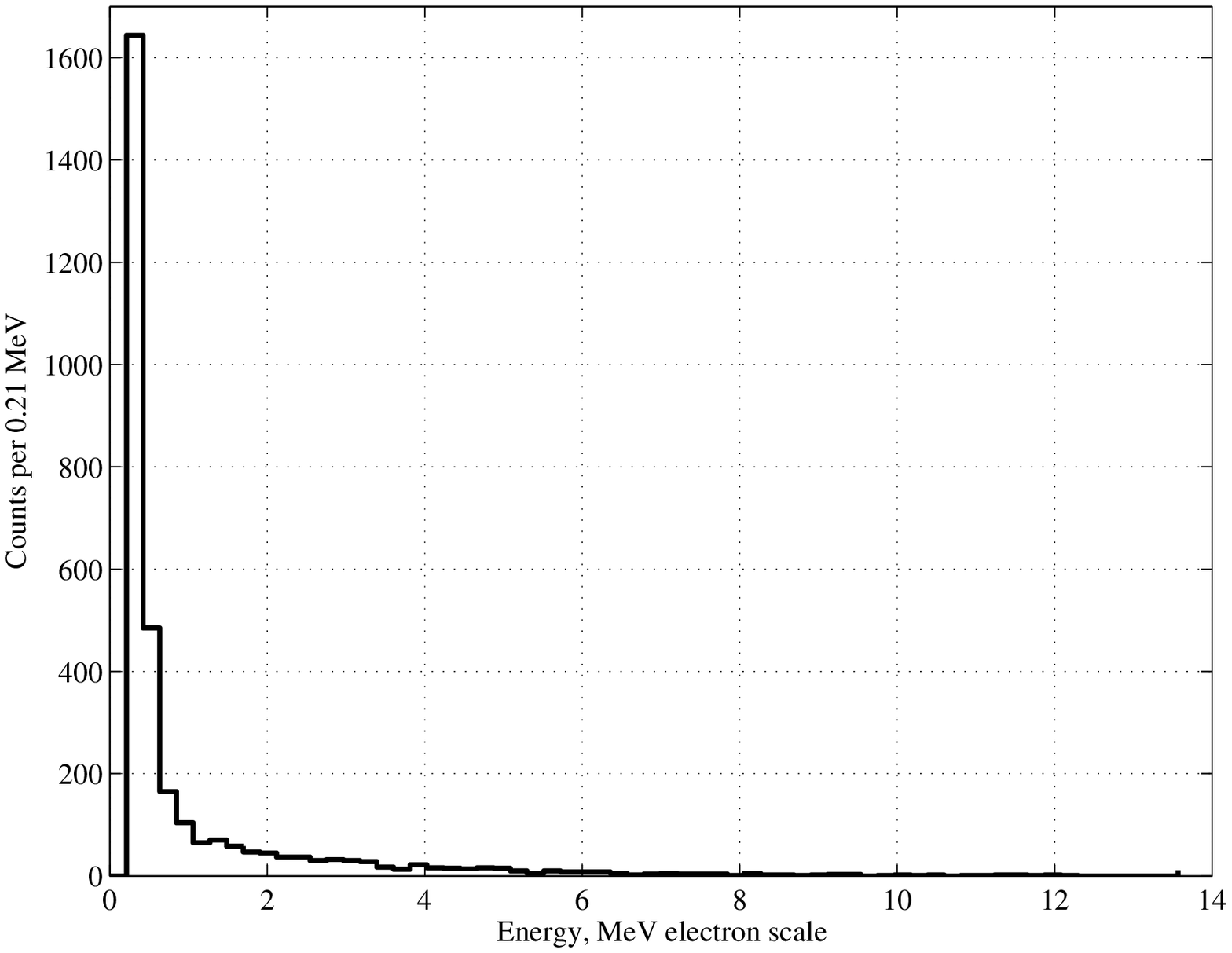}
\caption{The pulse height histogram of the PMT events correlated with the NC events in the neutron window at 660~m}
\label{pncor}
\end{figure}

Next, the simultaneous time window $STW$ has been checked for the correlated NC events in the energy region 70--170~keV (Figure~\ref{electr_del}). One founds the $STW=[-10;6]$, or 16~$\mu$s, the $CTW=[-160;-10]~\mu$s and the $ATW=[6;160]~\mu$s there. It's interesting to look at time delay distributions for the neutron window and $\alpha$-window helium counters events (Figure~\ref{neutr_del} and Figure~\ref{alfa_del}). Again, one can see clear fast neutron signature there. Then the PMT histograms were collected. Those histograms usually contain nothing specific --- see, for example, Figure~\ref{pncor}, where the PMT pulse height spectrum of the events coincident with the NC pulses in the neutron window is presented. For particular region (10--24 ADC~ch, or 1.6--3.0~MeV of neutron energy) one finds after 1.055 correction applied the $PNC=572\pm25$~events as a correlated signal. At the same region one finds the background $PNA+(PAC-PAA)=272\pm32$~events after the 1.055 and 2.164 corrections. Thus, the difference corrected by 1.38 and 1.1 is a pure neutron signal $FNS=454\pm62$, which results in the neutron rate in this energy region of $0.82\pm0.11$~h$^{-1}$. Then, using the efficiency $0.12\pm0.01$ and the surface 6267~cm$^2$ one finds the flux in the energy region of $[3.04\pm0.41]\times10^{-7}$~cm$^{-2}$s$^{-1}$. The same procedure is applied to other energy regions. If a negative value is obtained there after subtraction one uses the uncertainty as an upper limit for the flux.

Finally, one finds the slow neutron signal $SNS$. The integral of the fast neutron signal is $FNS=734\pm104$~events. Thus, one finds the $SNS=T_{nw}-T_{aw}-FNS=4151\pm319$~events. With the efficiency $0.16\pm0.03$ it results in the flux $[20.81\pm0.60]\times10^{-7}$ cm$^{-2}$s$^{-1}$ in the neutron energy region 0--1.6~MeV. The efficiency uncertainty isn't applied there.

\section{Results and discussion}

An average rate of fast neutron detection events deep underground is something like $\sim$1 h$^{-1}$ giving thus several hundreds events per month roughly. Due to low statistics we perform result of measurements in different energy ranges rather than spectrum plots.

Table~\ref{table4} contains results of the measurement of the neutron background at different levels of the Pyh\"asalmi mine. Uncertainties expressed explicitely includes both statistical and systematical ones. In addition, there are uncertainties of average detection efficiency for every energy region. The neutron energy above $\sim$1.5~MeV has been measured directly as an amplitude of light flash in the scintillator coincident with capture in $^3$He counters. Count rate of the coincident events is a base for flux calculation. Neutron with energy $<$1.5~MeV does not produce detectable light flash but can be effectively detected by helium counters. The neutron flux with those energies is deduced from count rate of helium counters subtracted by fast neutron rate. Those fluxes that appeared to be under uncertainties are expressed as an upper limits.

\begin{table*}[t]
\caption{ Neutron background at different levels of the Pyh\"asalmi mine, $10^{-7}$cm$^{-2}$s$^{-1}$.}\label{table4}
\begin{tabular}{lllllll}
\hline
Energy,&\multicolumn{6}{l}{Energy released region}\\
MeV&0--1.5&1.5--3&3--6&6--12&12--25&$>$25\\ \hline
Efficiency&0.16$\pm$0.03&0.12$\pm$0.01&0.10$\pm$0.01&0.08$\pm$0.01&0.05$\pm$0.02&0.01$\pm$0.01\\ 
%&&&&&&\\				
400 m&26.1$\pm$1.7&2.8$\pm$0.6&2.1$\pm$0.7&$<$0.8&$<$0.3&1.9$\pm$0.2\\
660 m&20.8$\pm$1.6&3.0$\pm$0.4&1.9$\pm$0.5&$<$0.7&$<$0.2&$<$0.4\\
990 m&37.5$\pm$1.7&6.2$\pm$0.6&2.6$\pm$0.7&$<$0.7&$<$0.9&$<$0.6\\
1410 m&42.2$\pm$5.0&10.5$\pm$2.4&3.0$\pm$1.9&3.3$\pm$1.5&$<$0.9&$<$0.7\\ \hline
\end{tabular}
\end{table*}

At 400 m level one can see significant excess in the energy region $>$25~MeV. An obvious reason is a production of neutrons in hadron showers generated by cosmic muons in the lead shielding. Muon induced neutrons mostly have an evaporating spectrum that is their energy lies in the range 1--10~MeV, and in principle can be measured in the detector. But in this case the energy released by evaporated neutrons in the scintillator is accompanied with and usually much less than the energy deposited by muon induced hadron showers. So, it can't be distinguished. Nevertheless, we put those neutrons in high energy released region just to separate from other ones.

Significant excess of neutrons at 1410~m can be easily reasoned by higher contamination of U/Th in the decorative granite powder covering the walls of the room where the measurement was done.

\section*{Acknowledgments}

We thank academician V.~Matveev, professor V.~Bezrukov for great interest to this work and intensive help. We acknowledge our colleagues from the CUPP project, B.~Brusila personally for providing us with excellent conditions of the work. Thank to head, management of the INMET~Co., T.~Myaki personally and workers of the Pyh\"asalmi mine for their understanding of needs of science and help in job. Many thanks to our colleagues from the INR RAS: G.~Abdullina, S.~Girin, N.~Gorshkov, T.~Ibragimova, A.~Kalikhov, I.~Mirmov and N.~Mirmova for the help and interesting discussions.

\end{document}